\documentclass[12pt, letterpaper]{article}

\usepackage[margin=2cm]{geometry}
\usepackage[utf8]{inputenc} 
\usepackage[T1]{fontenc}    
\usepackage{hyperref}       
\usepackage{url}            
\usepackage{booktabs}       
\usepackage{amsfonts}       
\usepackage{nicefrac}       
\usepackage{microtype}      
\usepackage{lipsum,todonotes,tikz}
\usepackage{cite}
\usepackage{graphicx,color}

\usepackage{mathtools}

\usepackage{physics}

\newcommand{\at}[2][]{#1\bigg|_{#2}}

\newcommand{\change}[1]{#1}
\usepackage{enumitem}
\usepackage{sectsty} 
\usepackage{graphicx, float}
\title{Entropic dynamics on Gibbs  statistical manifolds}

\author{Pedro Pessoa$^1$, Felipe Xavier Costa$^1$, Ariel Caticha$^1$  \\ $^1$Department of Physics, University at Albany - SUNY \\  Albany, NY, USA }
\date{}
\begin{document}
\maketitle
\abstract{
Entropic dynamics is a framework in which the laws of dynamics are derived as an application of entropic methods of inference. Its successes include the derivation of quantum mechanics and quantum field theory from probabilistic principles. Here we develop the entropic dynamics of a system the state of which is described by a probability distribution. Thus, the dynamics unfolds on a statistical manifold which is automatically endowed by a metric structure provided by information geometry. The curvature of the manifold has a significant influence. We focus our dynamics on the statistical manifold of Gibbs distributions (also known as canonical distributions or the exponential family). The model includes an \textquotedblleft entropic\textquotedblright\ notion of time that is tailored to the system under study; the system is its own clock. As one might expect, entropic time is intrinsically directional; there is a natural arrow of time which is lead by entropic considerations. As illustrative examples we discuss dynamics on a space of Gaussians and the discrete 3-state system.

}



\section{Introduction}
 
The original method of Maximum Entropy (MaxEnt) is usually associated with the names of Shannon \cite{Shannon-1948} and Jaynes \cite{Jaynes-1957I,Jaynes-1957II, Jaynes-1983, Jaynes-2003} although its roots can be traced to Gibbs \cite{Gibbs-1902}. The method was designed to assign probabilities on the basis of partial information in the form of expected value constraints and the central quantity, called entropy, which was interpreted as a measure of uncertainty or as an amount of missing information. In a series of developments starting with Shore and Johnson \cite{Shore-Johnson-1980}, with further contributions from other authors \cite{Skilling-1988, Caticha-2003,  Caticha-2007, Caticha-Giffin-2006, Vanslette-2017}, the range of applicability of the method was significantly extended. In its new incarnation the purpose of the method of Maximum Entropy, which will be referred as ME to distinguish it from the older version, is to update probabilities from arbitrary priors when new information in the form of constraints is considered \cite{Caticha-2020}. Highlights of the new method include: (1) A unified treatment of Bayesian and entropic methods which demonstrates their mutual consistency. (2) A new concept of entropy as a tool for reasoning that requires no interpretation in terms of heat, multiplicities, disorder, uncertainty, or amount of information. Indeed, entropy in ME needs no interpretation; it is a tool designed to perform a certain function -- to update probabilities to accommodate new information.   (3) A Bayesian concept of information defined in terms of its effects on the beliefs of rational agents --- the constraints are the information. (4) The possibility of information that is not in the form of expected value constraints. (We shall see an example below.)

The old MaxEnt was sufficiently versatile to provide the foundations to equilibrium statistical mechanics \cite{Jaynes-1957I} and to find application in a wide variety of fields such as economics \cite{Caticha-Golan-2014}, ecology \cite{Harte-2011, Jayanth-et-al-2010}, cellular biology \cite{De-Martino-et-al-2018, Dixit-et-al-2020}, network science \cite{Cimini19,Radicchi20}, and opinion dynamics \cite{Vicente-et-al-2014, Alves-et-al-2016}. As is the case with thermodynamics, all these applications are essentially static. MaxEnt has also been deployed to non-equilibrium statistical mechanics (see \cite{Jaynes-1979, Balian-1991} and subsequent literature in maximum caliber, e.g.\cite{Presse-2013, Gonzalez-2015, Cafaro-2016}) but the dynamics is not intrinsic to the probabilities; it is induced by the underlying Hamiltonian dynamics of the molecules. 
For problems beyond physics there is a need for more general dynamical frameworks based on information theory.

The ME version of the maximum entropy method offers the possibility of developing a true dynamics of probabilities. It is a dynamics driven by entropy --- an Entropic Dynamics (ED) ---  which is automatically consistent with the principles for updating probabilities. ED naturally leads to an \textquotedblleft entropic\textquotedblright\ notion of time. Entropic time is a device designed to keep track of the accumulation of changes. Its construction involves three ingredients: one must introduce the notion of an instant, verify that these instants are suitably ordered, and finally one must define a convenient notion of duration or interval between successive instants. A welcome feature is that entropic time is tailored to the system under study; the system is its own clock. Another welcome feature is that such an entropic time is intrinsically directional --- an arrow of time is generated automatically. 

ED has been successful in reconstructing dynamical models in physics such as quantum mechanics \cite{Caticha-2010, Caticha-2019}, quantum field theory \cite{Ipek-et-al-2019}, and the renormalization group \cite{Pessoa-et-al-2018}. Beyond physics, it has been recently applied to the fields of finance \cite{Abedi-et-al-2019a, Abedi-et-al-2019b} and neural networks \cite{NCaticha-2020}. 
Here we aim for a different class of applications of ED: to describe the dynamics of Gibbs distributions, also known as canonical distribution (exponential family) in statistical physics (statistics), since they are the distributions defined by a set of expected values constraint, namely sufficient statistics.  
Unlike the other cited papers on ED, here we will not focus on what the distributions are meant to represent. Other assumptions that would be specific to the modeled system are beyond the scope of the present article.

The goal is to study the ED generated by transitions from one distribution to another. 
The main assumptions are that changes happen and that they are not discontinuous. We do not explain why changes happen --- this is a mechanics without a mechanism. Our goal is to venture an educated estimate of what changes one expects to happen. The second assumption is that systems evolve along continuous trajectories in the space of probability distributions. It also implies that the study of motion involves two tasks. The first is to describe how a single infinitesimal step occurs. The second requires a scheme to keep track of how a large number of these short steps accumulate to produce a finite motion. It is the latter task that involves the introduction of the concept of time.

The fact that the space of macrostates is a statistical manifold --- each point in the space is a probability distribution --- has a profound effect on the dynamics. The reason is that statistical manifolds are naturally endowed with a Riemannian metric structure given by the Fisher-Rao information metric (FRIM) \cite{Fisher-1925, Rao-1945}, this structure is known as information geometry \cite{Amari-et-al-2000, Amari-2016, Ay-at-al-2017}. The particular case of Gibbs distributions leads to additional interesting geometrical properties (see e.g. \cite{Caticha-2015, Nielsen-et-al}) which have been explored in the extensive work relating statistical mechanics to information geometry \cite{Ruppeiner-1995, Janyszek-1990, Brody-et-al-1995, Oshima-1999, Brody-et-al-2008,Yapage-2008, Tanaka-2017, Nicholson-et-al-2018}. Information geometry has also been used as a fundamental concept for complexity measures \cite{Ay-2011,Felice14, Cafaro-2018}.

In this paper we tackle the more formal aspects of an ED on Gibbs manifolds and offer a couple of illustrative examples. The formalism is applied to two important sets of probability distributions: the space of Gaussians and the space of distributions for a 3-state system, both of which can be written in the exponential form. Since these distributions are both well-studied and scientifically relevant it can give us a good insight on how \change{the dynamics works.}

It is important to emphasize that Gibbs distributions are not restricted to the description of a system in thermal equilibrium. While it is true that if one chooses the conserved quantities in Hamiltonian motion as the sufficient statistics the resultant Gibbs distributions are the ones associated to equilibrium statistical mechanics,  Gibbs distribution can be defined for arbitrary choices of sufficient statistics, the modeling endeavour includes identifying the ones that are relevant to the problem at hand. On the same note, the dynamics developed here is not a form of nonequilibrium statistical mechanics, which is driven by a underlying physical molecular dynamics,  while ED is completely agnostic of any microstate dynamics. 

The article is organized as follows: the next section discusses the space of Gibbs distributions and its geometric properties; section 3 considers the ideas of ED; section 4 tackles the difficulties associated with formulating ED on the curved space of probability distributions; section 5 introduces the notion of entropic time; section 6 describes the evolution of the system in the form of a differential equation; in section 7 we offer two illustrative examples of ED on a Gaussian manifold and on a 2-simplex.

\section{The statistical manifold of Gibbs distributions}

\subsection{Gibbs distributions} 

The canonical or Gibbs probability distributions are the macrostates 
of a system. They describe a state of uncertainty about the microstate $x\in\mathcal{X}$ of the macroscopic system. A canonical distribution $\rho(x)$
is assigned by maximizing the entropy

\begin{equation}
S[\rho|q]=-\int dx\,\rho(x)\log\frac{\rho(x)}{q(x)}\label{maxent}%
\end{equation}
relative to the prior $q(x)$ subject to $n$ expected value constraints

\begin{equation}
\int dx\,\rho(x)a^{i}(x)=A^{i}, \quad\text{with}\quad i=1\ldots n~,\label{1b}%
\end{equation}
and the normalization of $\rho(x)$. Typically the prior $q(x)$ is chosen to be a
uniform distribution over the space $\mathcal{X}$ so that it is maximally
non-informative, but this is not strictly necessary. The $n$ constraints, on
the other hand, reflect the information that happens to be  relevant
to the problem. The resulting canonical distribution is

\begin{equation}
\rho(x|\lambda)=\frac{q(x)}{Z(\lambda)}\exp[-\lambda_{i}a^{i}(x)],\label{canonicaldefinition}%
\end{equation}
where $\lambda=\{\lambda_{1}\ldots\lambda_{n}\}$ are the Lagrange multipliers
associated to the expected value constraints, and we adopt the Einstein
summation convention. The normalization constant is

\begin{equation}
Z(\lambda)=\int dx\,q(x)\exp[-\lambda_{i}a^{i}(x)]=e^{-F(\lambda)},\label{3a}%
\end{equation}
where $F=-\log Z$ plays a role analogous to the free energy. The Lagrange
multipliers $\lambda_{i}(A)$ are implicitly defined by

\begin{equation}\label{3b}
\frac{\partial F}{\partial\lambda_{i}}=A^{i} \ .
\end{equation}
Evaluating the entropy (1) at its maximum yields 
\begin{equation} \label{thermalentropy}
S(A) = - \int dx \ \rho(x|\lambda(A)) \log\frac{ \rho(x|\lambda(A))}{q(x)}=  \lambda_i(A) A^i - F (\lambda(A)) \ .
\end{equation}
which we shall call the macrostate entropy or (when there is no risk of confusion) just the entropy. Equation \eqref{thermalentropy} shows that $S(A)$ is the Legendre transform of $F(\lambda)$: a small change $dA^{i}$ in the constraints shows that $S(A)$ is indeed a function of the expected values $A^{i}$,

\begin{equation}
dS=\lambda_{i}dA^{i}\quad\text{so that}\quad\lambda_{i}=\frac{\partial
S}{\partial A^{i}}~.\label{4b}%
\end{equation}  

One might think that defining a dynamics on the family of canonical distributions might be too restricted to be of interest; however this family has widespread applicability. Here it has been derived using the method of maximum entropy but historically it is also been known as the exponential family, namely the only family of distributions that possess sufficient statistics. {Interestingly, this was a problem proposed by Fisher \cite{Fisher-1922} in the primordium of statistics and later proved independently by Pitman \cite{Pitman-1936}, Darmois \cite{Darmois-1935}, and Koopman \cite{Koopman-1936}}. 
The sufficient statistics turn out to be the functions $a_{i}(x)$ in \eqref{maxent}. In \autoref{distro} we give a short list of the priors $q(x)$ and the functions $a_{i}(x)$ that lead to well-known distributions \cite{Nielsen-et-al,Brody-2007}. 

\begin{table}
\caption{\label{distro}Identification of sufficient statistics, priors and Lagrange multipliers for some well-known probability distributions.}
\footnotesize
\begin{tabular}{l|lll}
\toprule
\textbf{Distribution}  \ & $\lambda$ parameter & \ \textbf{Suff. Stat.}  & \textbf{Prior} \\
\midrule
Exponent Polynomial \  \  $ \rho(x|\beta) = \frac{\sqrt[k]{\beta}}{\Gamma(1+1/\beta)} e^{-\beta x^k} \ $ & $\lambda = \beta$ & $a(x) = x^k $   & uniform\\
Gaussian \ \  $ \rho(x|\mu,\sigma) = \frac{1}{\sqrt{2\pi\sigma^2}} \exp\left[ - \frac{(x-\mu)^2}{2\sigma^2} \right] \ \ $ & $\lambda = \left( -\frac{\mu}{\sigma^2} \ ,  \frac{1}{2\sigma^2} \right) $ & $a(x) = ( x, x^2 ) $    & uniform\\
Multinomial (k) \ \ $\rho(x|\theta) = \frac{n!}{x_1! \ldots x_k!}\theta_1^{x_1} \ldots \theta_k^{x_k} $ & $\lambda = -\log ( \theta_1, \theta_2, \ldots , \theta_k) \ $  & $a=(x_1, \ldots x_k) \ $ & $q(x) =   \prod^k_{i=1} x_i! $ \\
Poisson \ \ $\rho(x|m) = \frac{m^x}{x!} e^{-m} $& $\lambda = -\log m $ & $a(x) = x$ &$q(x) = 1/x!$\\
Mixed power laws \ \ $\rho(x|\alpha,\beta) = \frac{x^{-\alpha} e^{-\beta x}}{\beta^{\alpha-1} \Gamma(1-\alpha)} $ & $\lambda = (\alpha,\beta)$ & $a = ( \log x, x)$  & uniform \\
\bottomrule
\end{tabular}
\end{table}

Naturally, the method of maximum entropy assumes that the various constraints are compatible with each other so that the set of multipliers $\lambda$ exists. It is further assumed that the constraints reflect physically relevant
information so that the various functions such as $A^i(\lambda) = \pdv{}{\lambda_i} F $ and $\lambda_i(A) = \pdv{}{A^i} S$ that appear in
the formalism are both invertible and differentiable and so that the space of
Gibbs distributions is indeed a manifold. However, the manifold may include singularities
of various kinds which are of particular interest as they may describe phenomena
such as phase transitions \cite{Ruppeiner-1995,Brody-et-al-2008}.

\subsection{Information Geometry} 

In order to establish the notation and to recall some results that will be needed in later sections, we offer a brief review of well known results concerning the information geometry of Gibbs distributions \cite{Caticha-2015,Amari-2016}.

To each set of expected values $A=\{A^{1}, A^2, \ldots , A^{n}\}$, or to the associated
set of Lagrange multipliers $\lambda=\{\lambda_1, \lambda_2, \ldots , \lambda_{n}\}$, there
corresponds a canonical distribution. Therefore the set of distributions
$\rho(x|\lambda)$ or, equivalently $\rho(x|A)$, is a statistical manifold in
which each point can be labelled by the coordinates $\lambda$ or by $A$.
Whether we choose $\lambda$ or $A$ as coordinates is purely a matter of
convenience. The change of coordinates is implemented using

\begin{equation}
\frac{\partial A^{i}}{\partial\lambda_{k}}=-\frac{\partial^{2}\log Z}%
{\partial\lambda_{k}\partial\lambda_{i}}=A^{k}A^{i}-\langle a^{k}a^{i}%
\rangle~,\label{6a}%
\end{equation}
where we recognize the covariance tensor,

\begin{equation}
C^{ij}=\langle(a^{i}-A^{i})(a^{j}-A^{j})\rangle=-\frac{\partial A^{i}%
}{\partial\lambda_{j}}~.\label{6b}%
\end{equation}
Its inverse is given by

\begin{equation} \label{Smetric}
C_{jk}=-\frac{\partial\lambda_{j}}{\partial A^{k}}=-\frac{\partial^{2}%
S}{\partial A^{j}\partial A^{k}} \ ,
\end{equation}
\change{that means, the inverse covariant matrix $C_{ij}$ is the Hessian of negative entropy in \eqref{thermalentropy}. This implies}%

\begin{equation}
C^{ij}C_{jk}=\frac{\partial A^{i}}{\partial A^{k}}=\delta_{k}^{i}~.\label{6d}%
\end{equation}

Statistical manifolds are endowed with  an essentially unique
quantity to measure the extent to which two neighboring distributions $\rho(x|A)$ and
$\rho(x|A+dA)$ can be distinguished from each other. This measure of
distinguishability provides a statistical notion of distance which is given by
FRIM,
$ d\ell^2 = {g_{ij} dA^i dA^j} $ where

\begin{equation}
\label{firstmetric}
    g_{ij} = \int dx \ \rho(x|A) \frac{\partial\log \rho(x|A)}{\partial A^i}\frac{\partial\log \rho(x|A)}{\partial A^j}  .
\end{equation}
For a broader discussion on existence, derivation and consistency of this metric, as well as its properties, see \cite{Caticha-2015,Amari-2016,Ay-at-al-2017}. It suffices to say here that FRIM is the unique metric structure that is invariant under Markov embeddings \cite{Cencov-1981,Campbell-1986} and; therefore, is the only way to assigning a differential geometry structure that is in accordance to the grouping property of probability distributions. 

To calculate $g_{ij}$ for canonical distributions we use

\begin{equation} 
g_{ij}=\frac{\partial\lambda_{k}}{\partial A^{i}}\frac{\partial\lambda_{l}%
}{\partial A^{j}}\int dx\,\rho\,\frac{\partial\log\rho}{\partial\lambda_{k}%
}\,\frac{\partial\log\rho}{\partial\lambda_{l}}%
\end{equation}
and

\begin{equation} 
\frac{\partial\log\rho(x|A)}{\partial\lambda_{k}}=A^{k}-a^{k}(x)
\end{equation}
so that, using (\ref{6a}-\ref{6d}), we have

\begin{equation}
g_{ij}=C_{ik}C_{lj}C^{kl}=C_{ij} \ .
\end{equation}
Therefore the metric tensor $g_{ij}$ is the inverse of the covariance matrix
$C^{ij}$ which, by \eqref{Smetric}, is the Hessian of the entropy. 

As mentioned above, instead of $A^{i}$ we could use the Lagrange multipliers
$\lambda_{i}$ as coordinates. Then the information metric is the covariance matrix, 

\begin{equation}\label{covariant metric} 
g^{ij}=\int dx\,\rho(x|\lambda)\,\frac{\partial\log\rho(x|\lambda)}%
{\partial\lambda_{i}}\,\frac{\partial\log\rho(x|\lambda)}{\partial\lambda_{j}%
}=C^{ij} \ .
\end{equation}
Therefore the distance $d\ell$ between neighboring distributions can be
written in either of two equivalent forms,

\begin{equation}
d\ell^{2}=g_{ij}dA^{i}dA^{j}=g^{ij}d\lambda_{i}d\lambda_{j}~.\label{5d}%
\end{equation}

Incidentally, the availability of a unique measure of volume $dV=(\det
g_{ij})^{1/2}d^{n}A$ implies that there is a uniquely defined notion of the 
uniform distribution over the space of macrostates. The uniform distribution $P_u$ 
assigns equal probabilities to equal volumes, so that

\begin{equation}
P_u(A)d^{n}A\propto g^{1/2}d^{n}A\quad\text{where}\quad g=\det g_{ij}~.
\end{equation}

To conclude this overview section, we note that the metric tensor $g_{ij}$ can
be used to lower the contravariant indices of a vector to produce its dual
covector. Using \eqref{Smetric} and \eqref{firstmetric} the covector $dA_{i}$ dual to the infinitesimal
vector with components $dA^{i}$ is

\begin{equation}
dA_{i}=g_{ij}dA^{j}=-\frac{\partial\lambda_{i}}{\partial A^{j}}dA^{j}%
=-d\lambda_{i}~.
\end{equation}
which shows that not only are the coordinates $A$ and $\lambda$ related
through a Legendre transformation, which is a consequence of entropy
maximization, but also through a vector-covector duality, i.e. $-d\lambda_i$ is  the covector dual to $dA^i$, which is a consequence of information geometry.   

\section{Entropic Dynamics }

Having established the necessary background, we can now develop an entropic framework to describe dynamics on the space of macrostates.    

\subsection{Change happens}  

Our starting assumption is that changes happen continuously, which is supported by observation in nature. Therefore the dynamics we wish to formulate assumes that the
system evolves along continuous paths. This assumption of continuity
represents a significant simplification because it implies that a finite
motion can be analyzed as the accumulation of a large number of
infinitesimally short steps. Thus, our first goal will be to find the
probability $P(A^{\prime}|A)$ that the system takes a short step from the
macrostate $A$ to the neighboring macrostate $A^{\prime}=A+dA$. The transition
probability $P(A^{\prime}|A)$ will be assigned by maximizing an entropy. This
requires, first, that we identify the particular entropy that is relevant to
our problem. Next, we must decide on the prior distribution: what short steps
we might expect before we know the specifics of the motion. Finally, we
stipulate the constraints that are meant to capture the information that is
relevant to the particular problem at hand.

To settle the first item --- the choice of entropy --- we note that not only
we are uncertain about the macrostate at $A$ but we are also uncertain about the
microstates $x\in\mathcal{X}$. This means that the actual universe of
discourse is the joint space $\mathcal{A\times X}$ and the appropriate
statistical description of the system is in terms of the joint distribution

\begin{equation} \label{superstatistics}
P(x,A) = P(x|A)P(A) =\rho(x|A)P(A) \ ,
\end{equation}
 Where $\rho$ is of form  \eqref{canonicaldefinition} which means that we impose $P(x|A)$ to be canonical and the distribution $P(A)$ represents our lack of knowledge about the macrostates. { Note that what we did in \eqref{superstatistics} is nothing more than assuming a probability distribution for the macrostates. This description is sometimes referred to as superstatistics \cite{Beck-2003}. }

Our immediate task is to find the transition probability of a change
$P(x^{\prime},A^{\prime}|x,A)$ by maximizing the entropy%

\begin{equation}
\mathsf{S}[P|Q]=-\int dA^{\prime}dx^{\prime}P(x^{\prime},A^{\prime}%
|x,A)\log\frac{P(x^{\prime},A^{\prime}|x,A)}{Q(x^{\prime},A^{\prime}%
|x,A)} \ ,  \label{Stransition}%
\end{equation}
relative to the prior $Q(x^{\prime},A^{\prime}|x,A)$ and subject to
constraints to be discussed below. (To simplify the notation in
multidimensional integrals we write $d^{n}A^{\prime}=dA^{\prime}$ and
$d^{n}x^{\prime}=dx^{\prime}$.)

 Although  $S$ in \eqref{thermalentropy} and $\mathsf{S}$ in \eqref{Stransition}  are both entropies, in the information theory sense, they represent two very distinct statistical objects.
The $S(A)$ in \eqref{thermalentropy} is the entropy of the macrostate -- which is what one may be used to from statistical mechanics --, while the $\mathsf S[P|Q]$ in \eqref{Stransition} is the entropy  to be maximized so that we find the transition probability that better matches the information at hand, which means $\mathsf S$ is a tool to select the dynamics of the macrostates.

\subsection{The Prior}  

We adopt a prior that implements the idea that the system evolves by taking
short steps $A\rightarrow A+\Delta A$ at the macrostate level but is otherwise
maximally uninformative. We write  

\begin{equation}
Q(x^{\prime},A^{\prime}|x,A)=Q(x^{\prime}|x,A,A^{\prime})Q(A^{\prime}|x,A) \ ,
\end{equation}
and analyze the two factors in turn. We shall assume that a priori, before we
know the relation between the microstates $x$ and the macrostate $A$, the
prior distribution for $x^{\prime}$ is the same uniform underlying measure
$q(x^{\prime})$ introduced in \eqref{maxent},

\begin{equation}
Q(x^{\prime}|x,A,A^{\prime})=q(x^{\prime})~.
\end{equation}
Next we tackle the second factor $Q(A^{\prime}|x,A)$. As shown in appendix A, using the method of maximum entropy, the prior that enforces short steps but is
otherwise maximally uninformative is spherically symmetric as 

\begin{equation} 
Q(A^{\prime}|x,A)=Q(A^{\prime}|A) \propto g^{1/2}(A^{\prime})\exp[-\frac
{1}{2\tau}g_{ij}\Delta A^{i}\Delta A^{j}]~.
\end{equation}
so the joint prior is

\begin{equation} \label{priorfinal}
Q(x^{\prime},A^{\prime}|x,A) \propto q(x^{\prime}) g^{1/2}(A^{\prime})\exp[-\frac
{1}{2\tau}g_{ij}\Delta A^{i}\Delta A^{j}]~.
\end{equation}
We see that steps of length

\begin{equation}
\Delta\ell\sim(g_{ij}\Delta A^{i}\Delta A^{j})^{1/2} \gg \tau^{1/2} \ ,
\end{equation}
have negligible probability.  Eventually we will take the limit
$\tau\rightarrow0$ to enforce short steps. The prefactor $g^{1/2}(A^{\prime})$ ensures that
$Q(A^{\prime}|A)$ is a probability density.
Later, we will show how this choice of priors, that comes only from the assumption of continuous motion, leads to a diffusion structure.

\subsection{The constraints}  

The piece of information we wish to codify through the constraints is the
simple geometric idea that the dynamics remains confined to the statistical
manifold $\mathcal{A}$. This is implemented by writing  

\begin{equation}
P(x^{\prime},A^{\prime}|x,A)=P(x^{\prime}|x,A,A^{\prime})P(A^{\prime
}|x,A)\label{16}%
\end{equation}
and imposing that the distribution for $x^{\prime}$ is a canonical
distribution

\begin{equation} \label{28}
P(x^{\prime}|x,A,A^{\prime})=\rho(x^{\prime}|A^{\prime})\in\mathcal{A}%
~.
\end{equation}
This means that given $A^{\prime}$ the distribution of $x^{\prime}$ is
independent of the initial microstate $x$ and macrostate $A$. The second
factor in \eqref{16}, $P(A^{\prime}|x,A)$, is the transition probability we seek, leading to the constraint

\begin{equation}\label{constraintfinal}
P(x^{\prime},A^{\prime}|x,A)=\rho(x^{\prime}|A^{\prime}) P(A^{\prime}|x,A).
\end{equation}
We note that this constraint is not, \change{ as is usual in applications of the method of maximum entropy}, in the form of an expected value. \change{It may appear
from \eqref{constraintfinal} that the transition probability $P(A^{\prime }|x,A)$ will be
largely unaffected by the underlying space of microstates. To the contrary,
as we shall see below -- \eqref{21} and \eqref{tautransition} -- the macrostate dynamics turns
out to be dominated by the entropy of the microstate distribution $\rho(x^{\prime }|A^{\prime })$.}

Depending on the particular system under consideration, one could formulate
richer forms of dynamics by imposing additional constraints. To give just one
example, one could introduce some drift relative to the direction specified by
a covector $F_{i}$ by imposing a constraint of the form $\langle\Delta
A^{i}\rangle F_{i}=\kappa$ (see \cite{Caticha-2019, Ipek-et-al-2019}). In this paper however we shall limit ourselves to what is perhaps the simplest case, the minimal ED described by the single
constraint \eqref{constraintfinal}.

\subsection{Maximizing the entropy}  

Substituting \eqref{priorfinal} and \eqref{constraintfinal} into \eqref{Stransition} and rearranging we find

\begin{equation}
\mathsf{S}[P|Q]=\int dA^{\prime}\,P(A^{\prime}|x,A)\left[  -\log
\frac{P(A^{\prime}|x,A)}{Q(A^{\prime}|A)}+S(A^{\prime})\right]
\end{equation}
where $S(A^{\prime})$ is the macrostate entropy given in \eqref{thermalentropy}. Maximizing
$\mathsf{S}$ subject to normalization gives

\begin{align}
P(A^{\prime}|x,A)  & \propto Q(A^{\prime}|A)e^{S(A^{\prime})}\nonumber\\
& \propto g^{1/2}(A^{\prime})\exp[-\frac{1}{2\tau}g_{ij}\Delta A^{i}\Delta
A^{j}+S(A^{\prime})]~.\label{21}%
\end{align}
It is noteworthy that $P(A^{\prime}|x,A)$ turned out to be independent of $x$,
which is not surprising since neither the prior nor the constraints indicate
any correlation between $A^{\prime}$ and $x$.

Since the transition from $A$ to $A'$ has to be an arbitrarily small continuous change, we perform a linear approximation of $S$. This makes the exponential factor in \eqref{21} quadratic in $\Delta A$ as 

\begin{equation} \label{tautransition}
    P(A'|A) = \frac{g^{1/2} (A')}{\mathcal Z}   \exp\left[  \frac{\partial S}{\partial A^i} \Delta A^i- \frac{1}{2 \tau}  g_{ij} \Delta A^i \Delta A^j \right],
\end{equation}
where $e^{S(A)}$ was absorbed in the normalization factor $\mathcal{Z}$. This is the transition probability  found by maximizing the entropy \eqref{Stransition}.
However some mathematical  difficulties  arise from the fact that \eqref{tautransition} is defined over a curved manifold. 
We are going to explore these mathematical issues and their consequences to motion in the following section.

\section{The transition probability}
Since the statistical manifold is a curved space, we must understand how the transition probability \eqref{tautransition} behaves under a change of coordinates. 
As \eqref{priorfinal} and  \eqref{tautransition} describe an arbitrarily small step, we wish to express the transition probability, as well as quantities derived from it, calculated up to the order of  $\tau$. 
As the exponent in \eqref{tautransition} is manifestly invariant, so that one can complete squares and obtain 

\begin{equation} \label{nongaussiantransition}
    P(A'|A) = \frac{g^{1/2} (A')}{\mathcal Z'}   \exp\left[  - \frac{1}{2 \tau}  g_{ij}\left(  \Delta A^i - \tau g^{ik}\frac{\partial S}{\partial A^k} \right)  \left(  \Delta A^j   - \tau  g^{ik} \frac{\partial S}{\partial A^k}  \right) \right]  \  .
\end{equation}
If $g(A)$ were uniform, it would imply that the first two moments  $\expval{\Delta A^i}$ and $\expval{\Delta A^i \Delta A^j}$ are of order $\tau$. Therefore, even in the limit $\tau \rightarrow 0$, the transition will be affected by  curvature effects. This can be verified for an arbitrary metric tensor by a direct calculation of the first moment,

\begin{equation}\begin{split} \label{firstfirstmoment}
\langle \Delta A^i \rangle =& \int dA' \ \Delta A^i  P(A'|A)  \\
=& \frac{1}{\mathcal Z'}  \int dA'  \ g^{1/2} (A') \Delta A^i \exp\left[  - \frac{ g_{kl}}{2 \tau}\left(  \Delta A^k -  \tau V^k \right)  \left(  \Delta A^l   -  \tau V^l \right) \right] \ ,
\end{split}\end{equation}
where 
$    V^i =  g^{ij} \frac{\partial S}{\partial A^j} \ $. And the second moment

\begin{equation}\begin{split} \label{firstsecondmoment}
\langle \Delta A^i \Delta A^j\rangle =& \int dA' \ \Delta A^i  \Delta A^j  P(A'|A)  \\
=& \frac{1}{\mathcal Z'}  \int dA'  \ g^{1/2} (A') \Delta A^i \Delta A^j  \exp\left[    - \frac{ g_{kl}}{2 \tau}\left(  \Delta A^k -  \tau  V^k \right)  \left(  \Delta A^l   -  \tau  V^l \right) \right] \ .
\end{split}\end{equation}

To facilitate the calculation of the integrals in \eqref{firstfirstmoment} and \eqref{firstsecondmoment} it is convenient to write \eqref{tautransition} in normal coordinates at $A$. \change{Meaning, for a smooth manifold one can always make a change of coordinates $A^\mu(A^i)$ --  we will label the normal coordinates with Greek letter indexes ($\mu , \nu$) -- so that the metric in this coordinate system is so that}   
  
  \begin{equation}
      g_{\mu \nu}(A) = \delta_{\mu\nu}\  \ \text{and} \ \ \pdv{g_{\mu\nu}}{A^\mu}\at{A} = 0 \ ,
  \end{equation}
  allowing us to approximate $g(A') =1 $ for a short step. \change{For a general discussion and rigorous proof of the existence of normal coordinates see \cite{Kobayashi63}. Although normal coordinates are a valuable tool for geometrical analysis at this point, it is not clear whether they can be given a deeper statistical interpretation -- this is unlike other applications of differential geometry, such as general relativity, where the physical interpretation of normal coordinates turns out be of central importance.    }
  A displacement in these coordinates $\Delta A^{\mu}$ can be related to the original coordinates by a Taylor expansion in terms of $ \Delta A^{i}$ as (see \cite{Nawaz-2016,Nelson-1985}) 

\begin{equation}\label{taylor}
    \Delta A^{\mu} = \frac{\partial A^{\mu}}{\partial A^i} \Delta A^i +   \frac{1}{2} \frac{\partial^2 A^{\mu}}{\partial A^j\partial A^k} \Delta A^j \Delta A^k + o(\tau) \ .
\end{equation}
To proceed, it is interesting to recall the Christoffel symbols $\Gamma^i_{jk}$, 

\begin{equation}
    \Gamma^i_{jk} = \frac{1}{2} g^{il}\left( \pdv{g_{jl}}{A^l}+\pdv{g_{lj}}{A^k}-\pdv{g_{jk}}{A^l}\right) \ , 
\end{equation}
which transform as 

\begin{equation}
    \Gamma^i_{jk} = \frac{\partial A^i}{\partial A^\mu}  \frac{\partial A^\nu}{\partial A^j}  \frac{\partial A^\sigma}{\partial A^k} \Gamma^\mu_{\nu\sigma} - \frac{\partial A^i}{\partial A^\mu}  \frac{\partial^2 A^{\mu}}{\partial A^j\partial A^k}  \ . 
\end{equation}
Since in normal coordinates we have $\Gamma^\mu_{\nu\sigma} =0$, this allows us to isolate $\Delta A^i$ up to the order $\tau$ obtaining

\begin{equation} \label{transform}
    \Delta A^i = \frac{\partial A^i}{\partial A^{\mu}} \Delta A^{\mu} - \frac{1}{2} \Gamma^i_{jk}\Delta A^j \Delta A^k \ ,
\end{equation}
By squaring \eqref{transform} we have

\begin{equation}
    \label{transform2}
    \Delta A^i \Delta A^j = \frac{\partial A^i}{\partial A^{\mu}}\frac{\partial A^j}{\partial A^{\nu}}  \Delta A^\mu \Delta A^\nu +o(\tau) \ .
\end{equation}

Since the exponent in \eqref{firstfirstmoment} is invariant and in a coordinate transformation we have  $   dA \ P(A) =  d\tilde{A} \ P(\tilde{A}) $, it separates in two terms. 

\begin{equation}\begin{split} \label{secondfirstmoment}
\langle \Delta A^i \rangle &=   \frac{\partial A^i}{\partial A^{\mu}} \frac{1}{\mathcal Z'}  \int dA'  \Delta A^\mu \exp\left[  - \frac{ \delta_{\nu\sigma}}{2 \tau}\left(  \Delta A^\nu -  \tau  V^\nu \right)  \left(  \Delta A^\sigma   -  \tau  V^\sigma \right) \ \right] \\
&-\frac{1}{2} \Gamma^i_{jk}  
\frac{\partial A^\change{j}}{\partial A^{\mu}}\frac{\partial A^\change{k}}{\partial A^{\nu}}
\frac{1}{\mathcal Z'}  \int dA'  \Delta A^\mu \Delta A^\nu \exp\left[  - \frac{ \delta_{\upsilon\sigma}}{2 \tau} \left(  \Delta A^\nu -  \tau  V^\nu \right)  \left(  \Delta A^\sigma   -  \tau  V^\sigma \right) \right] \ .
\end{split}\end{equation}

The integrals can be evaluated from the known properties of a Gaussian. The integral in the fist term gives $\langle  \Delta A^\mu \rangle = \tau \delta^{\mu\nu} \pdv{S}{A^\nu}$  and the integral in the second term gives $\langle  \Delta A^\mu \change{\Delta}A^\nu \rangle = \tau \delta^{\mu\nu} $ so that

\begin{equation}
    \langle \Delta A^i \rangle = \frac{\partial A^i}{\partial A^{\mu}} \tau \delta^{\mu\nu} \pdv{S}{A^\mu}  -\frac{1}{2} \Gamma^i_{jk} 
    \frac{\partial A^\change{j}}{\partial A^{\mu}}\frac{\partial A^\change{k}}{\partial A^{\nu}}
    \tau \delta^{\mu\nu} \ .
\end{equation}
Therefore in natural coordinates the first two moments up to order of $\tau$ are

\begin{equation} \label{moments}
    \langle \Delta A^{i} \rangle = \tau g^{ij} \frac{\partial  S}{\partial A^j} - \frac{  \tau}{2} \Gamma^i \ , \ \ \text{and} \ \   \langle \Delta A^{i}\Delta A^{j} \rangle  = \tau g^{ij}  \ ,
\end{equation}
where $\Gamma^i = \Gamma^i_{jk} g^{jk}$. 
\change{Here we see the dependence on curvature for $\langle \Delta A^{i} \rangle$ in the Christoffel symbol term. Note that it is a consequence of the dependance between $\Delta A^{i}$ and the quadratic term $\Delta A^{i} \Delta A^{j}$ in \eqref{transform}, which per \eqref{moments} does not vanish even for small steps. Hence fluctuations in $A^i$ cannot be ignored in the ED motion and this is the reason why the motion probes curvature. It also follows from \eqref{moments} that even in the limit $\tau \rightarrow 0$ the average change $\Delta A^{i}$ does not transform covariantly. }

Note that we used several words such as \textquotedblleft transitions\textquotedblright\, \textquotedblleft short step\textquotedblright\, \textquotedblleft continuous\textquotedblright\ and \textquotedblleft dynamics\textquotedblright\ without any established notion of time. 
In the following section we will discuss time not as an external parameter, but as an emergent parameter from the maximum entropy transition \eqref{tautransition} and its moments \eqref{moments}.

    \section{Entropic time}
    
    Having described a short step transition the next challenge is to study how these short steps accumulate.

    \subsection{Introducing time}  

    In order to introduce time we note that $A'$ and $A$ are elements of the same manifold, therefore  $P(A')$ and $P(A)$ are two probability distributions over the same space.
    Our established solution to describe the accumulation of changes (see \cite{Caticha-2010}) is to introduce a \textquotedblleft book-keeping\textquotedblright\  parameter $t$ that distinguishes the said distributions as labelled by different parameters, i.e.    $P(A') = P_{t'}(A)$ and $P(A) = P_{t}(A)$.  
    
    In this formalism we will refer to these different labels as a description of the system at particular instants $t$ and $t'$. This allow us to call $P(A'|A)$ a transition probability.
    
    \begin{equation}\label{CK}
        P_{t'}(A) = P(A') =  \int dA \ P_{\Delta t}(A'|A) P_t(A)
    \end{equation} 
    where $\Delta t = t' - t$.
    
    As the system changes from $A$ to $A'$ and then to $A''$. The probability $P(A'')$ will be constructed from $P(A')$, not explicitly dependent on $P(A)$.  This means that \eqref{CK} represents a Markovian process: conditioned on the present $P_{t'}(A)$, the \textquotedblleft future\textquotedblright\ $P_{t''}(A)$ is independent of the \textquotedblleft past\textquotedblright\ $P_{t}(A)$, where $t'' > t'>t$.
    It is important to notice that, under this formalism, \eqref{CK} is not used to show that the  process is Markovian in an existing time, but rather the concept of time  developed here  makes the dynamics Markovian by design.
    
    It is also important to notice that the parameter $t$ presented here is not necessarily the \textquotedblleft physical\textquotedblright\ time (as it appears in Newton's laws of motion or the Schr\"{o}dinger equation). 
    Our parameter $t$, which we call entropic time, is an epistemic well-ordered parameter in which the dynamics is defined. 
    
   \subsection{The entropic arrow of time}  
    It is important to note that the marginalization process \change{from \eqref{superstatistics} to \eqref{CK}} could also lead to
    
    \begin{equation}
    P(A) = \int dA' \ P(A|A') P(A') \ ,
    \end{equation}
     where the conditional probabilities are related by Bayes' Theorem,
     
    \begin{equation}
    \label{Bayes}
    P(A|A') = \frac{P(A)}{P(A')} P(A'|A)  \ ,
    \end{equation}
    showing that a change \textquotedblleft forward\textquotedblright\  will not happen the same way as a change \textquotedblleft backwards\textquotedblright\  unless the system is in some form of stationary state, $P(A)=P(A')$.
     Another way to present this is that probability theory alone gives no intrinsic distinction of the change \textquotedblleft forward\textquotedblright\  and \textquotedblleft backward\textquotedblright\ . The fact that we assigned the change \textquotedblleft forward\textquotedblright\  by ME implies that, in general, the change \textquotedblleft backward\textquotedblright\  is not an entropy maximum.
     Therefore, the preferential direction of the flow of time arises from the entropic dynamics naturally.

    \subsection{Calibrating the clock}  

In order to connect the entropic time to the transition probability, one needs to define the duration $\Delta t$ with respect to the motion. 
Time in entropic dynamics is defined so as to simplify the description of the motion. This notion of time is tailored to the system under discussion.
The time interval will be chosen so that the parameter $\tau$ that first appeared in the prior \eqref{priorfinal} takes the role of a time interval, 

\begin{equation} \label{clock}
\tau =  \eta \Delta t \ ,
\end{equation}
where $\eta$ is a constant so that $t$ has the units of time. For the remainder of this article we will adopt $\eta = 1$. 
In principle any monotonic function $t(\tau)$ serves as an  parameter for ordering. Our choice is a matter of convenience, as required by simplicity. 
Here this is implemented so that for a short transition we have the dimensionless time interval

\begin{equation}\label{transformwfinal}
     \Delta t = {g_{ij}} \langle \Delta A^{i}\Delta A^{j} \rangle \ .   
\end{equation}
This means that the entropic time is  measured by the system's fluctuations. Rather than having the changes in the system  represented in terms of given time intervals (as measured by an external clock), here the system is its own clock.

The moments in \eqref{moments} can be written, up to order $\Delta t$, as

\begin{equation} \label{moments2}
    \frac{\langle \Delta A^{i} \rangle}{\Delta t} =  g^{ij} \frac{\partial  S}{\partial A^j} - \frac{  1 }{2} \Gamma^i \ , \ \ \text{and} \ \  \frac{ \langle \Delta A^{i}\Delta A^{j} \rangle}{\Delta t}  =   g^{ij}  \ .
\end{equation}
With this we have  established a concept of time, it is convenient to write the trajectory of the expected values in terms of a differential equation.

\section{Diffusion and the Fokker-Planck equation}

Our goal of designing the dynamics from entropic methods is accomplished. The entropic dynamics equation of evolution is written in integral form as a Chapman-Kolmogorov equation  \eqref{CK} with a transition probability given by \eqref{tautransition}. In this section we will conveniently rewrite it in the differential form.
The computed drift $\langle \Delta A^i \rangle$  and the fluctuation $\langle \Delta A^i\Delta A^j \rangle$ in \eqref{moments2} describe the dynamical process as a smooth diffusion --{ meaning, as defined by \cite{Nelson-1985}, a stochastic process in which the first two moments are, calculated to the order of $\Delta t$,  $\langle\Delta A^i \rangle= b^i \Delta t$, $\langle\Delta A^{i} \Delta A^{j} \rangle=  \eta g^{ij} \Delta t$ and $\langle\Delta A^{i} \Delta A^{j} \Delta A^{k} \rangle=0$}. Therefore, for a short transition,  it is possible to write the evolution of $P_t(A)$,
as a Fokker-Planck (diffusion) equation,

 \begin{equation} \label{FP1}
\frac{\partial}{\partial t}P=-\frac{\partial}{\partial A^i} \left(  P v^{i}\right)  \ , 
\end{equation}
where

\begin{equation} \label{FP1vel}
    v^i =  g^{ij} \frac{\partial S}{\partial A^j} - \frac{1}{2} g^{ij} \frac{\partial}{\partial A^j}\left(\log \frac{ P}{g^{1/2}}\right) \ .
\end{equation}

The  derivation of \eqref{FP1}  and \eqref{FP1vel} takes into account the fact that the space in which the diffusion happens is curved and is given in appendix B.
In  equation \eqref{FP1vel} we see that the current velocity $v^i$ consists of two components. The first term is the drift velocity guided by the entropy gradient and the second term is an osmotic velocity, that is a term that is driven by differences in probability density. The examples presented in the following section  will show how these terms interact and the dynamical properties derived from each.

\subsection{Derivatives and divergence}\label{li}  

Since  the entropy $S$ is a scalar,  the velocity defined in \eqref{FP1vel} is a contravariant vector. However, \eqref{FP1} is not a manifestly invariant equation. To check its consistency it is convenient to write it in terms of the invariant object $p$ defined as

\begin{equation} \label{pscalar}
    p(A) = \frac{P(A)}{g^{1/2}(A)}  \ ,
\end{equation}
\change{meaning $p$ is the probability of $A$ divided by the volume element,} in terms of which  \eqref{FP1} becomes

\begin{equation}\label{FP2}
     \frac{\partial}{\partial t}p =- \frac{1}{g^{1/2}} \frac{\partial}{\partial A^i} \left( g^{1/2}\ p  v^{i}\right)  \ .
\end{equation}
We can recognize, on the right-hand side, the covariant divergence of the contravariant vector $p v^i$, which can be written in the manifestly covariant form

\begin{equation} \label{FPinv}
     \frac{\partial}{\partial t}p =- \mathcal{D}_i ( p v^i ) \ ,
\end{equation} 
where $\mathcal{D}_i$ is the covariant derivative. 
\change{The fact that the covariant derivative arises from the dynamical process is the direct indication that even when evolving the invariant object $p$  the curvature of the space is taken into account.}
We can identify \eqref{FPinv} as a continuity equation   \change{-- generalized to the parallel transport in a curved space, as evidenced by the covariant divergence --} where the flux, $j^i = p v^i$, can be written from  \eqref{FP1vel} and \eqref{pscalar} as

\begin{equation}
    j^i = p g^{ij} \pdv{S}{A^j} - \frac{1}{2} g^{ij} \pdv{p}{A^j} \ .
\end{equation}
The second term, which is related to the osmotic velocity, is a Fick's law with diffusion tensor $D^{ij}= g^{ij}/2 $. Note that this is identified from purely probabilistic arguments, rather than assuming a repulsive interaction from the microstate dynamics.

Having the dynamics fully described we can now study its consequences as it will be done in the following section.

\section{Examples} \label{examples}
We established the entropic dynamics by finding the transition probability \eqref{tautransition}, presenting it as a differential equation in \eqref{FP1}, \eqref{FP1vel} and presenting it as the invariant equation \eqref{FPinv}.
We want to show some examples on how it would be applied and what are the results achieved. Our present goal is not to search for realistic models, but to search for models which are both mathematically simple and general enough so it can give insight on how to use the formalism.

We will be particularly interested in two properties: the drift velocity, 

\begin{equation}\label{eq:vd}
    v_D^i =  g^{ij}\pdv{S}{A^j} \ ,
\end{equation} which is the first term in \eqref{FP1vel}, and the  static states, $v^i=0$, which are a particular subset of the dynamical system's equilibrium $\partial_t P = 0$. Obtained from   \eqref{FP1vel} as

\begin{equation}
v^i = 0 \Rightarrow \frac{\partial S}{\partial A^i} - \frac{1}{2} \frac{\partial}{\partial{A^i}} \log(\frac{P}{g^{1/2}}) = 0
\end{equation}
allowing one to write the static probability

\begin{equation}\label{eq:stationary}
P(A) \propto g^{1/2}(A) \exp[2S(A)] \ ,
\end{equation}
where the factor of $2$ in the exponent comes from the diffusion tensor $D^{ij} = g^{ij}/2$ explained in section \ref{li}.  
This result shows that the invariant stationary probability density \eqref{pscalar} is

\begin{equation}\label{eq:stationaryDensity}
p(A) \propto \exp[2S(A)].
\end{equation}

\subsection{A Gaussian Manifold}  

The statistical manifold defined by the mean values and correlations of a random variable, the microstate $x$, is the space of Gaussian distribution, which is an example of a canonical distribution. Here we consider the dynamics of a two-dimensional spherically symmetric Gaussian with a non-uniform variance, $\sigma(A) = \sigma(A^1, A^2)$, defined by

\begin{equation}
\expval{x^1} = A^{1}, \qq{} \expval{x^{2}} = A^{2}, \qq{and} \expval{ (x^{i}-A^{i})(x^{j}-A^{j}) } = \sigma^{2}(A)\delta^{ij}.
\end{equation}
These Gaussians are of the form,

\begin{equation}
\rho(x \mid A) = \frac{1}{2\pi\sigma^2(A)} \exp(- \frac1{2\sigma^2(A)} \sum_{i=1}^2 (x^i - A^i)^2) \label{Eq:Gauss}
\end{equation}
The entropy of \eqref{Eq:Gauss} relative to a uniform background measure is given by

\begin{equation}
S(A) = \log(2\pi \sigma(A)^2) \label{Eq:GaussEntropy}
\end{equation}

The space of Gaussians with a uniform variance, $\sigma(A)=\,$constant, is
flat and the dynamics turns out to be a rather trivial spherically symmetric
diffusion. Choosing the variance to be non-uniform yields a richer and more
interesting dynamics. Since this example is pursued for purely illustrative
purposes we restrict to two dimensions and to spherically symmetric Gaussians.
The generalization is immediate.


The FRIM for a Gaussian distribution is found, using \eqref{firstmetric} (see also \cite{Caticha-2020}), to be 

\begin{equation}
d l^2 = \frac{4}{\sigma^2} (d \sigma)^2 + \frac{\delta_{ij}}{\sigma^2} d A^i d A^j \ , \label{eq:GaussDist}
\end{equation}
so that, using

\begin{equation}
d \sigma = \pdv{\sigma}{A^i} d A^i~,
\end{equation}
the induced metric $d l^2 = g_{ij} d A^i d A^j$ leads to,

\begin{equation} \label{eq:inducedm}
g_{ij} = \frac1{\sigma^2} \left( 4 \pdv{\sigma}{A^i}\pdv{\sigma}{A^j} + \delta_{ij}\right) \ .
\end{equation}

\subsubsection{Gaussian submanifold around an entropy maximum}

We present an example of our dynamical model that illustrates the motion around an entropy maximum. 
A simple way to manifest it  is to recognize that, in \eqref{FP1vel}, $-S$ plays a role analogous to a potential. A rotationally symmetric quadratic potential can then be sustituted in \eqref{Eq:GaussEntropy} leading to 

\begin{equation}
\sigma(A) = \exp(-\frac{(A^1)^2 + (A^2)^2}4) \ ,
\end{equation}
which substituted in \eqref{eq:inducedm} yields the metric 

\begin{equation}\label{Gcurv}
g_{ij} = \mqty[\qty(A^1)^2 + \sigma^{-2} & A^1A^2 \\[1ex] A^1A^2 & \qty(A^2)^2 + \sigma^{-2} ] \ ,
\end{equation}
so that 

\begin{equation}
g^{1/2} = \sqrt{ \left[ \qty(A^1)^2 + \qty(A^2)^2\right]\sigma^{-2} + \sigma^{-4} } \ .
\label{eq:Uniform}
\end{equation}
\change{The scalar curvature for the Gaussian submanifold can be calculated from \eqref{Gcurv} as} 

\begin{equation}
    R = \frac{\phi^2-2\phi}{(\phi^2 + \sigma^{-2})^2} \sigma^2, \qq{where} \ \phi = {(A^1)^2 + (A^2)^2} \ .
\end{equation}

\begin{figure}[H]
  \begin{minipage}[b]{0.4\textwidth}
    \includegraphics[width=\textwidth]{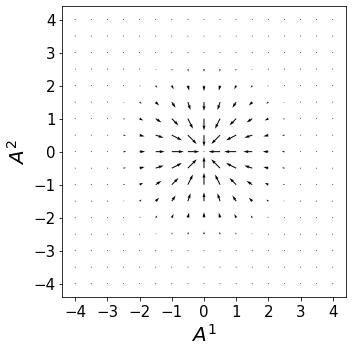}
    \caption{Drift velocity field \eqref{eq:DriftGaussMax} drives the flux along the entropy gradient.} 
    \label{fig:DriftGaussMax}
  \end{minipage}
  \hfill
  \begin{minipage}[b]{0.5\textwidth}
    \includegraphics[width=\textwidth]{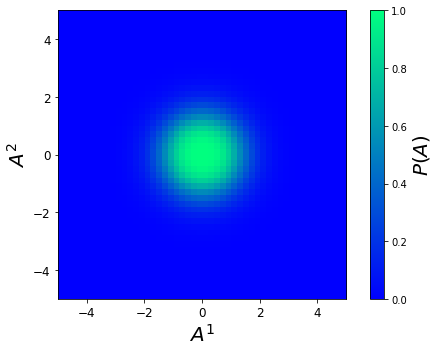}
    \caption{Equilibrium stationary probability \eqref{eq:PoGaussMax}}
    \label{fig:PoGaussMax}
  \end{minipage}
\end{figure}

\change{From \eqref{eq:vd} }
the drift velocity (figure~\ref{fig:DriftGaussMax}) is

\begin{equation}
\change{v_D^1} = -\frac{A^1 \sigma^{-2}}{g}  \qq{and} \change{v_D^2} = -\frac{A^2 \sigma^{-2}}{g} \ . \label{eq:DriftGaussMax}
\end{equation}
and, \change{from \eqref{eq:stationary}}, the static probability (figure~\ref{fig:PoGaussMax}) is

\begin{equation}
P(A) \propto 4\pi^2 g^{1/2} \sigma^{4} \label{eq:PoGaussMax} \ .
\end{equation}

The static distribution results from the dynamical equilibrium between two opposite tendencies. One is the drift velocity field that drives the distribution along the entropy gradient towards the entropy maximum at the origin. The other is the osmotic diffusive force that we identified earlier as the ED analogue of Fick's law. This osmotic force drives the distribution against the direction of the probability gradient and prevents it from becoming infinitely concentrated at the origin. At equilibrium the cancellation between these two opposing forces results in the Gaussian distribution, equation~\eqref{eq:PoGaussMax}.

\subsection{2-Simplex Manifold}  

Here we discuss an example of discrete microstates. The macrostate coordinates, being expected values, are continuous variables. Our subject matter will be a three-state system, $x \in \{1,2,3\}$, such as, for example, a 3-sided die. The statistical manifold is the 2-dimensional simplex and the natural coordinates are the probabilities themselves,

\begin{equation}
\mathcal{S}_{2} = \qty{\rho(x) \mid \rho(x) \geq 0 ~,~ \sum_{x=1}^3 \rho(x) = 1}.
\end{equation}

The distributions on the 2-simplex are Gibbs distributions defined by the
sufficient statistics of functions

\begin{equation}\label{eq:suffstat}
a^{i}(x)=\delta_{x}^{i}\quad\text{so that}\quad A^{i}=\langle a^{i}%
\rangle=\rho(i)~.
\end{equation}

\begin{figure}[H]
  \begin{minipage}[b]{0.45\textwidth}
    \includegraphics[width=\textwidth]{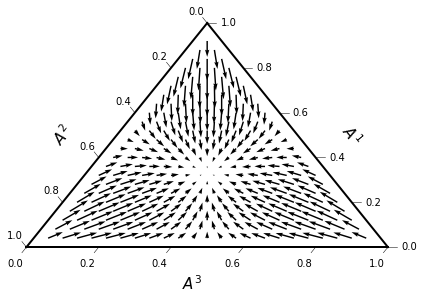}
    \caption{Drift velocity field for the 2-simplex in~\eqref{eq:Drift3die}. {The ternary plots ware created using python-ternary library \cite{pythonternary}.} }
    \label{fig:Drift3die}
  \end{minipage}
  \hfill
  \begin{minipage}[b]{0.5\textwidth}
    \includegraphics[width=\textwidth]{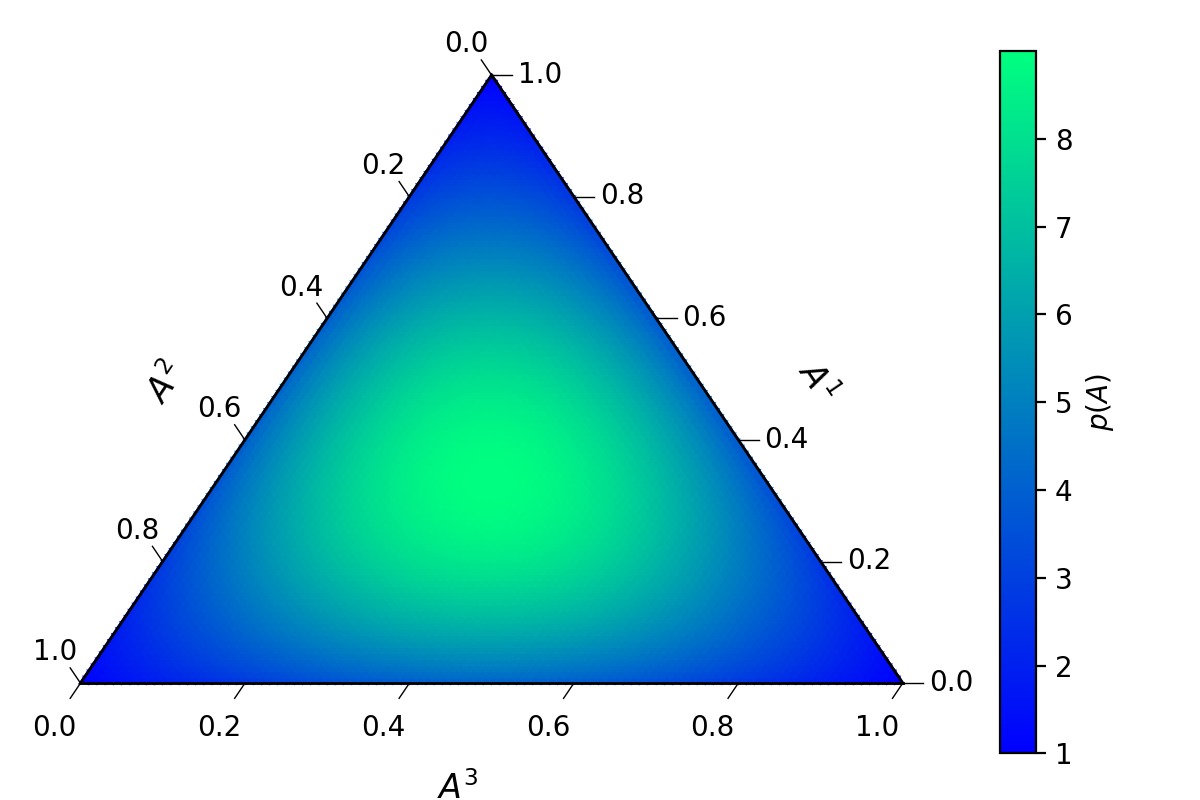}
    \caption{Static invariant stationary probability for the three-state system.}
    \label{fig:Po3die}
  \end{minipage}
\end{figure}

The entropy relative to the uniform discrete measure is

\begin{equation}
S = -\sum_{i=1}^{3} \rho(i) \log(\rho(i)) = - \sum_{i=1}^{3} A^i \log(A^i) \ ,
\end{equation}
and the information metric is given by

\begin{equation}
g_{ij} = \sum_{k=1}^{3} \rho^k \pdv{\log(\rho^k)}{A^i}\pdv{\log(\rho^k)}{A^j}~.
\end{equation}
The 2-simplex arises naturally from probability theory due to normalization when one identifies the macrostate of interest to be the probabilities themselves. 
The choice of sufficient statistics \eqref{eq:suffstat} implies that the manifold is a two-dimensional surface since, due to the normalization, one can write $A^3 = 1- A^1-A^2$. We will use the the tuple $(A^1,A^2)$ as our coordinates and $A^3$ as a function of them. In this scenario, one finds a metric tensor

\begin{equation}\label{2Smetric}
g_{ij} = \mqty[\dfrac1{A^3} + \dfrac{1}{A^1} & \dfrac1{A^3} \\[2ex] \dfrac1{A^3} & \dfrac1{A^3} + \dfrac1{A^2}]~,
\end{equation}
which induces the volume element

\begin{equation}
g^{1/2} =\sqrt{ \frac1{A^1 A^2 A^3} }~.
\end{equation}
\change{As is well known, the simplex is characterized by a constant curvature $R=1/2$; the
2-simplex is the positive octant of a sphere.  }
From \eqref{eq:vd} the drift velocity (figure~\ref{fig:Drift3die}) is

\begin{equation}
\begin{split}
    v^1_\change{D}  &= A^1\left[A^2\log(\frac{A^2}{A^3}) + (A^1-1)\log(\frac{A^1}{A^3})\right] \\
    v^2_\change{D} &= A^2\left[A^1\log(\frac{A^1}{A^3}) + (A^2-1)\log(\frac{A^2}{A^3})\right] \ ,
\end{split}\label{eq:Drift3die}
\end{equation}
Also, the static probability is

\begin{equation}
P(A) \propto g^{1/2}   \prod_{i=1}^3 \qty(A^i)^{-2A^i} \ .
\label{eq:Po3die}
\end{equation}

 We note from the determinant of the metric that the static probability~\eqref{eq:Po3die} diverges at the boundary of the 2-simplex. This reflects the fact that a 2-state system (say, $i=1,2$) is easily distinguishable from a 3-state system ($i=1,2,3$). Indeed, a single datum $i=3$ will tell us that we are dealing with a 3-state system.

On the other hand, we can see (fig.~\ref{fig:Po3die}) that this divergence is not present in the invariant stationary probability \eqref{pscalar}.

As in the Gaussian case discussed in the previous section, the static equilibrium results from the cancellation of two opposing forces: the entropic force along the drift velocity field towards the center of the simplex is cancelled by the osmotic diffusive force away from the center.

\section{Conclusions}

We conclude with a summary of the main results. In this paper the entropic dynamics framework has been extended to describe dynamics on a statistical
manifold.  ME played an instrumental role in that it allowed us to impose constraints that not are in the standard form of expected values.

The resulting dynamics, which follows from purely entropic considerations,
takes the form of a diffusive process on a curved space. The effects of
curvature turn out to be significant. We found that the probability flux is
the resultant of two components. One describes a flux along the entropy
gradient and the other is a diffusive or osmotic component that turns out to
be the curved-space analogue of Fick's law with a diffusion tensor
$D^{ij}=g^{ij}/2$ given by information geometry. 

A highlight of the model is that it includes an \textquotedblleft
entropic\textquotedblright\ notion of time that is tailored to the system
under study; the system is its own clock. This opens the door to the
introduction of a notion of time that transcends physics and might be useful
for social and ecological systems. The emerging notion of entropic time is
intrinsically directional. There is a natural arrow of time which manifests
itself in a simple description of the approach to equilibrium.

The model developed here is rather minimal in the sense that the dynamics
could be extended by taking additional relevant information into account. For
example, it is rather straightforward to enrich the dynamics by imposing
additional constraints

\begin{equation}
\langle\Delta A^{i}\rangle F_{i}(A)=\kappa^{\prime} \ ,
\end{equation}
involving system-specific functions $F_{i}(A)$ that carry information about
correlations. This is the kind of further developments we envisage in
future work. 

As illustrative examples, the dynamics was applied to two general spaces of probability distributions. A submanifold of the space of two-dimensional Gaussians and the space of probability distributions for a 3-state system (2-simplex). In each of these we were able to provide insight on the dynamics by presenting the drift velocity \eqref{eq:vd} and the equilibrium stationary states \eqref{eq:stationary}. Also, as future work, we intent to apply the dynamics developed here in the distributions found in network sciences \cite{Costa21}.

\appendix

\section{Obtaining the prior}

In this appendix we derive the prior transition probability from $A$  to $A'$ seen in \eqref{priorfinal}. This is achieved by maximizing the entropy

\begin{equation}\label{terceiraentropia}
    \text{S}[Q] = -\int dA' \ Q(A'|x,A) \log \left( \frac{Q(A'|x,A)}{R(A'|x,A)}\right) \ ,
\end{equation}
where $R(A'|x,A)$, the prior for \eqref{terceiraentropia}, encodes information about an unbiased transition of the systems. The posterior of \eqref{terceiraentropia}, $Q(A'|x,A)$, becomes the prior in \eqref{Stransition}.

At this stage $A$ could evolve into any $A'$ and the only assumption is that the assigned prior for \eqref{terceiraentropia} leads to equal probabilities for equal volumes; thus, ignoring biases.
That is achieved by a prior proportional to the volume element $R(A'|x,A) \ \propto \ g^{1/2}(A')$, where $g(A) = \det g_{ij}(A)$. There is no need to address the normalization of $R$ since it will have no effect on the posterior.

The chosen constraint represents an isotropic and continuous motion on the manifold. This will be imposed by

\begin{equation} \label{appendixconstraint}
    \int dA' \ Q(A'|x,A) \ g_{ij} \Delta A^i \Delta A^j = K \ .
\end{equation}
where $K$ is a small quantity, since $g_{ij} \Delta A^i \Delta A^j$ is invariant only in the limit for short steps $\Delta A^i \to 0$. Therefore, eventually, $K \to 0$.

The result of maximizing \eqref{terceiraentropia} under \eqref{appendixconstraint} and normalization is

\begin{equation}
    Q(A'|x,A) \propto  g^{1/2} (A') \exp\left(  - \alpha \ g_{ij} \Delta A^i \Delta A^j \right) \ ,
\end{equation}
where $\alpha$ is the Lagrange multiplier associated to \eqref{appendixconstraint}. As the result requires $K \rightarrow 0$ to make it geometrically invariant, the conjugated Lagrange multiplier should be allowed to be taken to infinity. This allows us to define $\tau = 1/\alpha$, such that the short step limit will lead to $\tau \to 0$.

Note that, since no motion in $x$ and no correlation between $x$ and $A'$ is induced by the constraints, the result does not depend on the previous microstate $x$, $Q(A'|x,A) = Q(A'|A)$.

\section{Derivation of the Fokker-Planck equation}\label{FPApp}

The goal of this appendix is to show that for a dynamics that is a smooth diffusion in a curved space, can be written as a Fokker-Planck equation and to obtain its velocity \eqref{FP1vel} from the moments for the motion \eqref{moments2}.
In order to do so, it is convenient to define a drift velocity

\begin{equation} \label{b}
    b^i = \lim_{\Delta t \rightarrow 0}  \frac{\langle \Delta A^{i} \rangle}{\Delta t} = g^{ij} \frac{\partial S}{\partial A^j} - \frac{1}{2} \Gamma^i \ .
\end{equation}

First, let us analyze the change of a smooth integrable function $f(A)$ as the system transitions from $A$ to $A'$. A smooth change in the function $f(A)$ will be

\begin{equation} \label{diferencetestfunction}
    \Delta f(A)  = \frac{\partial f}{\partial A^i} \Delta A^i + \frac{1 }{2} \frac{\partial^2 f}{\partial A^i\partial A^j} \Delta A^i\Delta A^j + o(\Delta t) \ ,
\end{equation}
since a cubic term, $\Delta A^i\Delta A^j\Delta A^k $ would be $o(\Delta t)$. In a smooth diffusion we can take the expected value of \eqref{diferencetestfunction} with respect to $P(A'| A)$ as

\begin{equation}
\label{averagetestfunction}
    \langle\Delta f(A)  \rangle = \int dA' \ P(A'|A)  (f(A')-f(A)) = \left( b^i \frac{\partial }{ \partial A^i } + \frac{1}{2} g^{ij} \frac{\partial^2 }{\partial A^i\partial A^j}  \right) f(A) \Delta t \ . 
\end{equation}
which can be further averaged in $P(A)$. The left-hand side will be

\begin{align}
\int dA \  P(A) \int dA' \ P(A'|A)  (f(A')-f(A)) &= \int dA dA' P(A', A) (f(A')-f(A)) \\
&= \int dA' \  P(A') f(A') -  \int dA \  P(A) f(A)
\end{align}
while the right hand is

\begin{equation}
\int dA \ P(A) \left( b^i \frac{\partial }{ \partial A^i } + \frac{1}{2} g^{ij} \frac{\partial^2 }{\partial A^i\partial A^j}  \right) f(A) \Delta t \ . 
\end{equation}
such that they equate to 

\begin{equation}
\int dA' \  P(A') f(A') -  \int dA \  P(A) f(A) = \int dA \ P(A) \left( b^i \frac{\partial }{ \partial A^i } + \frac{1}{2} g^{ij} \frac{\partial^2 }{\partial A^i\partial A^j}  \right) f(A) \Delta t \ . 
\end{equation}

As established in section 5,  $P(A)$ and $P(A') $ are distributions at the instants $t$ and $t'$ respectively. 

\begin{equation}
     \int dA \ \left(\frac{ P_{t'}(A) - P_{t}(A)}{\Delta t}\right) f(A )  
    = \int dA \ P(A) \left( b^i \frac{\partial }{ \partial A^i } + \frac{1}{2} g^{ij} \frac{\partial^2 }{\partial A^i\partial A^j}  \right) f(A) \Delta t \ ,
\end{equation}
which can be partially integrated in the limit of small steps

\begin{equation}
     \int dA \ 
    \left( \pdv{P(A)}{t} \right) f(A)= 
    \int dA \  \left( -  \frac{\partial}{ \partial A^i } (b^i P(A)) + \frac{1}{2}  \frac{\partial^2 }{\partial A^i\partial A^j} ( g^{ij}P(A))  \right) f(A) \ .
\end{equation}
Due to the generality of  $f$ as test function, we identify the integrants,  

\begin{equation}\label{fourterms}
    \frac{\partial}{\partial t} P(A) = - \frac{\partial}{ \partial A^i }\left(    b^i P(A)  - \frac{1}{2} \frac{\partial }{\partial A^j} (  g^{ij} P(A))  \right) \ ,
\end{equation}
and substitute $b^i$ \eqref{b} for general coordinates,

\begin{equation}
    \frac{\partial}{\partial t} P(A) = - \frac{\partial}{ \partial A^i }\left(     g^{ij} \frac{\partial S}{\partial A^j} P(A)  - \frac{1 }{2} \Gamma^i P(A) - \frac{1 }{2} \left(\frac{\partial   g^{ij}}{\partial A^j} \right) P(A)    - \frac{1}{2} g^{ij} \frac{\partial P(A)}{ \partial A^j }\right) \ ,
\end{equation}
and the contracted Christoffel symbols can be substituted in the identity

\begin{equation} \label{magiclandauidentity}
\Gamma^i = - \frac{1}{g^{1/2}} \frac{\partial}{\partial A^j} (g^{1/2} g^{ij} )=  - \frac{\partial g^{ij}}{\partial A^j} - g^{ij} \frac{\partial \log g^{1/2}}{\partial A^j}  \ . 
\end{equation}

\change{Here we see that the effect of curvature -- encoded by the  Christoffel symbols  -- substitute in the differential equation \eqref{fourterms} obtaining}

\begin{equation}
    \frac{\partial}{\partial t} P(A) = - \frac{\partial}{ \partial A^i }\left(     g^{ij} \frac{\partial S}{\partial A^j}   - \frac{1}{2} g^{ij} \frac{\partial}{\partial A^j}\left(\log \frac{ P(A)}{g^{1/2}}\right) \right)P(A) \ ,
\end{equation}
\change{where the second term inside the parenthesis above is the result of taking the curvature into account.}
The result is a Fokker-Planck equation that is usefully written in the continuity form

 \begin{equation} \label{FP1a}
\frac{\partial}{\partial t}P=-\frac{\partial}{\partial A^i} \left(  P v^{i}\right)  \ , 
\end{equation}
where

\begin{equation} \label{FP1vela}
    v^i =  g^{ij} \frac{\partial S}{\partial A^j} - \frac{1}{2} g^{ij} \frac{\partial}{\partial A^j}\left(\log \frac{ P}{g^{1/2}}\right) \ ,
\end{equation}
completing the derivation.


%
%
%
%

\bibliographystyle{elsarticle-num}
\bibliography{Reference}

\begin{thebibliography}{10}
\expandafter\ifx\csname url\endcsname\relax
  \def\url#1{\texttt{#1}}\fi
\expandafter\ifx\csname urlprefix\endcsname\relax\def\urlprefix{URL }\fi
\expandafter\ifx\csname href\endcsname\relax
  \def\href#1#2{#2} \def\path#1{#1}\fi

\bibitem{Shannon-1948}
C.~E. Shannon, A mathematical theory of communication, The Bell System
  Technical Journal 27~(3) (1948) 379--423.
\newblock \href {https://doi.org/10.1002/j.1538-7305.1948.tb01338.x}
  {\path{doi:10.1002/j.1538-7305.1948.tb01338.x}}.

\bibitem{Jaynes-1957I}
E.~T. Jaynes, Information theory and statistical mechanics: I, Physical Review
  106~(4) (1957) 620.
\newblock \href {https://doi.org/10.1103/PhysRev.106.620}
  {\path{doi:10.1103/PhysRev.106.620}}.

\bibitem{Jaynes-1957II}
E.~T. Jaynes, Information theory and statistical mechanics. {II}, Physical
  Review 108~(2) (1957) 171.
\newblock \href {https://doi.org/10.1103/PhysRev.108.171}
  {\path{doi:10.1103/PhysRev.108.171}}.

\bibitem{Jaynes-1983}
R.~D. Rosenkrantz (Ed.), {E. T. Jaynes: Papers on Probability, Statistics and
  Statistical physics}, Reidel, Dordrecht, 1983.
\newblock \href {https://doi.org/10.1007/978-94-009-6581-2}
  {\path{doi:10.1007/978-94-009-6581-2}}.

\bibitem{Jaynes-2003}
E.~T. Jaynes, Probability theory: The logic of science, Cambridge University
  Press, 2003.

\bibitem{Gibbs-1902}
J.~Gibbs, {Elementary Principles in Statistical Mechanics}, Yale University
  Press, New Haven, 1902, .\ reprinted by Ox Bow Press, Connecticut 1981.

\bibitem{Shore-Johnson-1980}
J.~Shore, R.~Johnson, Axiomatic derivation of the principle of maximum entropy
  and the principle of minimum cross-entropy, IEEE Transactions on information
  theory 26~(1) (1980) 26--37.
\newblock \href {https://doi.org/10.1109/TIT.1980.1056144}
  {\path{doi:10.1109/TIT.1980.1056144}}.

\bibitem{Skilling-1988}
J.~Skilling, {The Axioms of Maximum Entropy}, in: G.~J. Erickson, C.~R. Smith
  (Eds.), {Maximum-Entropy and Bayesian Methods in Science and Engineering},
  Vol. 31-32, Springer, Dordrecht, 1988, pp. 173--187.
\newblock \href {https://doi.org/10.1007/978-94-009-3049-0_8}
  {\path{doi:10.1007/978-94-009-3049-0_8}}.

\bibitem{Caticha-2003}
A.~Caticha, \href{arXiv.org/abs/physics/0311093}{{Relative Entropy and
  Inductive Inference}}, in: {AIP Conference Proceedings}, Vol. 707, American
  Institute of Physics, 2004, pp. 75--96.
\newblock \href {https://doi.org/10.1063/1.1751358}
  {\path{doi:10.1063/1.1751358}}.
\newline\urlprefix\url{arXiv.org/abs/physics/0311093}

\bibitem{Caticha-2007}
A.~Caticha, \href{arXiv.org/abs/0710.1068}{{Information and Entropy}}, in: AIP
  Conference Proceedings, Vol. 954, American Institute of Physics, 2007, pp.
  11--22.
\newblock \href {https://doi.org/10.1063/1.2821253}
  {\path{doi:10.1063/1.2821253}}.
\newline\urlprefix\url{arXiv.org/abs/0710.1068}

\bibitem{Caticha-Giffin-2006}
A.~Caticha, A.~Giffin, \href{arxiv.org/abs/physics/0608185}{{Updating
  Probabilities}}, in: AIP Conference Proceedings, Vol. 872, American Institute
  of Physics, 2006, pp. 31--42.
\newblock \href {https://doi.org/10.1063/1.2423258}
  {\path{doi:10.1063/1.2423258}}.
\newline\urlprefix\url{arxiv.org/abs/physics/0608185}

\bibitem{Vanslette-2017}
K.~Vanslette, Entropic updating of probabilities and density matrices, Entropy
  19~(12) (2017) 664.
\newblock \href {https://doi.org/10.3390/e19120664}
  {\path{doi:10.3390/e19120664}}.

\bibitem{Caticha-2020}
A.~Caticha,
  \href{https://www.albany.edu/physics/faculty/ariel-caticha}{{Entropic
  Physics: Probability, Entropy, and the Foundations of Physics}}, 2012.
\newline\urlprefix\url{https://www.albany.edu/physics/faculty/ariel-caticha}

\bibitem{Caticha-Golan-2014}
A.~Caticha, A.~Golan, An entropic framework for modeling economies, Physica A:
  Statistical Mechanics and its Applications 408 (2014) 149--163.
\newblock \href {https://doi.org/10.1016/j.physa.2014.04.016}
  {\path{doi:10.1016/j.physa.2014.04.016}}.

\bibitem{Harte-2011}
J.~Harte, Maximum entropy and ecology: a theory of abundance, distribution, and
  energetics, OUP Oxford, 2011.

\bibitem{Jayanth-et-al-2010}
J.~R. Banavar, A.~Maritan, I.~Volkov, Applications of the principle of maximum
  entropy: from physics to ecology, Journal of Physics: Condensed Matter 22~(6)
  (2010) 063101.
\newblock \href {https://doi.org/10.1088/0953-8984/22/6/063101}
  {\path{doi:10.1088/0953-8984/22/6/063101}}.

\bibitem{De-Martino-et-al-2018}
A.~De~Martino, D.~De~Martino, An introduction to the maximum entropy approach
  and its application to inference problems in biology, Heliyon 4~(4) (2018)
  e00596.
\newblock \href {https://doi.org/10.1016/j.heliyon.2018.e00596}
  {\path{doi:10.1016/j.heliyon.2018.e00596}}.

\bibitem{Dixit-et-al-2020}
P.~D. Dixit, E.~Lyashenko, M.~Niepel, D.~Vitkup, Maximum entropy framework for
  predictive inference of cell population heterogeneity and responses in
  signaling networks, Cell Systems 10~(2) (2020) 204--212.
\newblock \href {https://doi.org/10.1101/137513} {\path{doi:10.1101/137513}}.

\bibitem{Cimini19}
G.~Cimini, T.~Squartini, F.~Saracco, D.~Garlaschelli, A.~Gabrielli,
  G.~Caldarelli, \href{https://doi.org/10.1038/s42254-018-0002-6}{The
  statistical physics of real-world networks}, Nature Reviews Physics 1~(1)
  (2019) 58--71.
\newblock \href {https://doi.org/10.1038/s42254-018-0002-6}
  {\path{doi:10.1038/s42254-018-0002-6}}.
\newline\urlprefix\url{https://doi.org/10.1038/s42254-018-0002-6}

\bibitem{Radicchi20}
F.~Radicchi, D.~Krioukov, H.~Hartle, G.~Bianconi,
  \href{https://doi.org/10.1088/2632-072x/ab9447}{Classical information theory
  of networks}, Journal of Physics: Complexity 1~(2) (2020) 025001.
\newblock \href {https://doi.org/10.1088/2632-072x/ab9447}
  {\path{doi:10.1088/2632-072x/ab9447}}.
\newline\urlprefix\url{https://doi.org/10.1088/2632-072x/ab9447}

\bibitem{Vicente-et-al-2014}
R.~Vicente, A.~Susemihl, J.~P. Jeric{\'o}, N.~Caticha, Moral foundations in an
  interacting neural networks society: A statistical mechanics analysis,
  Physica A: Statistical Mechanics and its Applications 400 (2014) 124--138.
\newblock \href {https://doi.org/10.1016/j.physa.2014.01.013}
  {\path{doi:10.1016/j.physa.2014.01.013}}.

\bibitem{Alves-et-al-2016}
F.~Alves, N.~Caticha, Sympatric multiculturalism in opinion models, in: AIP
  Conference Proceedings, Vol. 1757, AIP Publishing LLC, 2016, p. 060005.
\newblock \href {https://doi.org/10.1063/1.4959064}
  {\path{doi:10.1063/1.4959064}}.

\bibitem{Jaynes-1979}
E.~T. Jaynes, Where do we stand on maximum entropy?, in: R.~D. Levine,
  M.~Tribus (Eds.), The Maximum Entropy Principle, MIT Press, 1979, reprinted
  in [Jaynes 1983] and at http://bayes.wustl.edu.
\newblock \href {https://doi.org/10.1007/978-94-009-6581-2_10}
  {\path{doi:10.1007/978-94-009-6581-2_10}}.

\bibitem{Balian-1991}
R.~Balian, From Microphysics to Macrophysics: Methods and Applications of
  Statistical Mechanics. Volumes I and II, Springer Heidelberg, 1991 and 1992.

\bibitem{Presse-2013}
S.~Pressé, K.~Ghosh, J.~Lee, K.~A. Dill, Principles of maximum entropy and
  maximum caliber in statistical physics, Reviews of Modern Physics 85~(3)
  (2013) 1115–1141.
\newblock \href {https://doi.org/10.1103/revmodphys.85.1115}
  {\path{doi:10.1103/revmodphys.85.1115}}.

\bibitem{Gonzalez-2015}
S.~Davis, D.~González, Hamiltonian formalism and path entropy maximization,
  Journal of Physics A: Mathematical and Theoretical 48~(42) (2015) 425003.
\newblock \href {https://doi.org/10.1088/1751-8113/48/42/425003}
  {\path{doi:10.1088/1751-8113/48/42/425003}}.

\bibitem{Cafaro-2016}
C.~Cafaro, S.~A. Ali, Maximum caliber inference and the stochastic ising model,
  Physical Review E 94~(5) (2016).
\newblock \href {https://doi.org/10.1103/physreve.94.052145}
  {\path{doi:10.1103/physreve.94.052145}}.

\bibitem{Caticha-2010}
A.~Caticha, \href{arXiv.org/abs/1005.2357}{Entropic dynamics, time and quantum
  theory}, Journal of Physics A: Mathematical and Theoretical 44~(22) (2011)
  225303.
\newblock \href {https://doi.org/10.1088/1751-8113/44/22/225303}
  {\path{doi:10.1088/1751-8113/44/22/225303}}.
\newline\urlprefix\url{arXiv.org/abs/1005.2357}

\bibitem{Caticha-2019}
A.~Caticha, \href{arXiv.org/abs/1908.04693}{The entropic dynamics approach to
  quantum mechanics}, Entropy 21~(10) (2019) 943.
\newblock \href {https://doi.org/10.3390/e21100943}
  {\path{doi:10.3390/e21100943}}.
\newline\urlprefix\url{arXiv.org/abs/1908.04693}

\bibitem{Ipek-et-al-2019}
S.~Ipek, M.~Abedi, A.~Caticha, \href{arXiv.org/abs/1803.07493}{Entropic
  dynamics: reconstructing quantum field theory in curved space-time},
  Classical and Quantum Gravity 36~(20) (2019) 205013.
\newblock \href {https://doi.org/10.1088/1361-6382/ab436c}
  {\path{doi:10.1088/1361-6382/ab436c}}.
\newline\urlprefix\url{arXiv.org/abs/1803.07493}

\bibitem{Pessoa-et-al-2018}
P.~Pessoa, A.~Caticha, \href{arXiv.org/abs/1712.02267}{Exact renormalization
  groups as a form of entropic dynamics}, Entropy 20~(1) (2018) 25.
\newblock \href {https://doi.org/10.3390/e20010025}
  {\path{doi:10.3390/e20010025}}.
\newline\urlprefix\url{arXiv.org/abs/1712.02267}

\bibitem{Abedi-et-al-2019a}
M.~Abedi, D.~Bartolomeo, Entropic dynamics of exchange rates and options,
  Entropy 21~(6) (2019) 586.
\newblock \href {https://doi.org/10.3390/e21060586}
  {\path{doi:10.3390/e21060586}}.

\bibitem{Abedi-et-al-2019b}
M.~Abedi, D.~Bartolomeo, Entropic dynamics of stocks and european options,
  Entropy 21~(8) (2019) 765.
\newblock \href {https://doi.org/10.3390/e21080765}
  {\path{doi:10.3390/e21080765}}.

\bibitem{NCaticha-2020}
N.~Caticha, Entropic dynamics in neural networks, the renormalization group and
  the hamilton-jacobi-bellman equation, Entropy 22~(5) (2020) 587.
\newblock \href {https://doi.org/10.3390/e22050587}
  {\path{doi:10.3390/e22050587}}.

\bibitem{Fisher-1925}
R.~A. Fisher, Theory of statistical estimation, in: Proc. Cambridge Philos.
  Soc., Vol. 122, 1925, p. 700.
\newblock \href {https://doi.org/10.1017/S0305004100009580}
  {\path{doi:10.1017/S0305004100009580}}.

\bibitem{Rao-1945}
C.~R. Rao, Information and the accuracy attainable in the estimation of
  statistical parameters, in: Bull. Calcutta Math. Soc., Vol.~37, 1945, p.~81.
\newblock \href {https://doi.org/10.1007/978-1-4612-0919-5_16}
  {\path{doi:10.1007/978-1-4612-0919-5_16}}.

\bibitem{Amari-et-al-2000}
S.~Amari, H.~Nagaoka, Methods of information geometry, American Mathematical
  Soc., 2000.

\bibitem{Amari-2016}
S.~Amari, Information geometry and its applications, Springer, 2016.
\newblock \href {https://doi.org/10.1007/978-4-431-55978-8}
  {\path{doi:10.1007/978-4-431-55978-8}}.

\bibitem{Ay-at-al-2017}
N.~Ay, J.~Jost, H.~V. L{\^{e}}, L.~Schwachh\"{o}fer,
  \href{https://doi.org/10.1007/978-3-319-56478-4}{Information Geometry},
  Springer International Publishing, 2017.
\newblock \href {https://doi.org/10.1007/978-3-319-56478-4}
  {\path{doi:10.1007/978-3-319-56478-4}}.
\newline\urlprefix\url{https://doi.org/10.1007/978-3-319-56478-4}

\bibitem{Caticha-2015}
A.~Caticha, \href{arXiv.org:1412.5633}{The basics of information geometry}, in:
  AIP Conference Proceedings, Vol. 1641, American Institute of Physics, 2015,
  pp. 15--26.
\newblock \href {https://doi.org/10.1063/1.4905960}
  {\path{doi:10.1063/1.4905960}}.
\newline\urlprefix\url{arXiv.org:1412.5633}

\bibitem{Nielsen-et-al}
F.~{Nielsen}, V.~{Garcia}, {Statistical exponential families: A digest with
  flash cards}, arXiv e-prints (2009) ,\href {http://arxiv.org/abs/0911.4863}
  {\path{arXiv:0911.4863}}.

\bibitem{Ruppeiner-1995}
G.~Ruppeiner, Riemannian geometry in thermodynamic fluctuation theory, Reviews
  of Modern Physics 67~(3) (1995) 605.
\newblock \href {https://doi.org/10.1103/RevModPhys.67.605}
  {\path{doi:10.1103/RevModPhys.67.605}}.

\bibitem{Janyszek-1990}
H.~Janyszek, R.~Mrugala, Riemannian geometry and stability of ideal quantum
  gases, Journal of Physics A: Mathematical and General 23~(4) (1990) 467.
\newblock \href {https://doi.org/0.1088/0305-4470/23/4/016}
  {\path{doi:0.1088/0305-4470/23/4/016}}.

\bibitem{Brody-et-al-1995}
D.~Brody, N.~Rivier, Geometrical aspects of statistical mechanics, Physical
  Review E 51~(2) (1995) 1006.
\newblock \href {https://doi.org/10.1103/PhysRevE.51.1006}
  {\path{doi:10.1103/PhysRevE.51.1006}}.

\bibitem{Oshima-1999}
H.~Oshima, T.~Obata, H.~Hara, Riemann scalar curvature of ideal quantum gases
  obeying gentiles statistics, Journal of Physics A: Mathematical and General
  32~(36) (1999) 6373–6383.
\newblock \href {https://doi.org/10.1088/0305-4470/32/36/302}
  {\path{doi:10.1088/0305-4470/32/36/302}}.

\bibitem{Brody-et-al-2008}
D.~Brody, D.~W. Hook, Information geometry in vapour--liquid equilibrium,
  Journal of Physics A: Mathematical and Theoretical 42~(2) (2008) 023001.
\newblock \href {https://doi.org/10.1088/1751-8113/42/2/023001}
  {\path{doi:10.1088/1751-8113/42/2/023001}}.

\bibitem{Yapage-2008}
N.~Yapage, H.~Nagaoka, An information geometrical approach to the mean-field
  approximation for quantum ising spin models, Journal of Physics A:
  Mathematical and Theoretical 41~(6) (2008) 065005.
\newblock \href {https://doi.org/10.1088/1751-8113/41/6/065005}
  {\path{doi:10.1088/1751-8113/41/6/065005}}.

\bibitem{Tanaka-2017}
S.~Tanaka, Information geometrical characterization of the onsager-machlup
  process, Chemical Physics Letters 689 (2017) 152--155.
\newblock \href {https://doi.org/10.1016/j.cplett.2017.10.005}
  {\path{doi:10.1016/j.cplett.2017.10.005}}.

\bibitem{Nicholson-et-al-2018}
S.~B. Nicholson, A.~del Campo, J.~R. Green, Nonequilibrium uncertainty
  principle from information geometry, Physical Review E 98~(3) (2018) 032106.
\newblock \href {https://doi.org/10.1103/PhysRevE.98.032106}
  {\path{doi:10.1103/PhysRevE.98.032106}}.

\bibitem{Ay-2011}
N.~Ay, E.~Olbrich, N.~Bertschinger, J.~Jost, A geometric approach to
  complexity, Chaos: An Interdisciplinary Journal of Nonlinear Science 21~(3)
  (2011) 037103.
\newblock \href {https://doi.org/10.1063/1.3638446}
  {\path{doi:10.1063/1.3638446}}.

\bibitem{Felice14}
D.~Felice, S.~Mancini, M.~Pettini, Quantifying networks complexity from
  information geometry viewpoint, Journal of Mathematical Physics 55~(4) (2014)
  043505.
\newblock \href {https://doi.org/10.1063/1.4870616}
  {\path{doi:10.1063/1.4870616}}.

\bibitem{Cafaro-2018}
D.~Felice, C.~Cafaro, S.~Mancini, Information geometric methods for complexity,
  Chaos: An Interdisciplinary Journal of Nonlinear Science 28~(3) (2018)
  032101.
\newblock \href {https://doi.org/10.1063/1.5018926}
  {\path{doi:10.1063/1.5018926}}.

\bibitem{Fisher-1922}
R.~A. Fisher, On the mathematical foundations of theoretical statistics,
  Philosophical Transactions of the Royal Society of London. Series A,
  Containing Papers of a Mathematical or Physical Character 222~(594-604)
  (1922) 309--368.
\newblock \href {https://doi.org/10.1098/rsta.1922.0009}
  {\path{doi:10.1098/rsta.1922.0009}}.

\bibitem{Pitman-1936}
E.~J.~G. Pitman, Sufficient statistics and intrinsic accuracy, in: Mathematical
  Proceedings of the cambridge Philosophical society, Vol.~32, Cambridge
  University Press, 1936, pp. 567--579.
\newblock \href {https://doi.org/10.1017/S0305004100019307}
  {\path{doi:10.1017/S0305004100019307}}.

\bibitem{Darmois-1935}
G.~Darmois, Sur les lois de probabilit{\'e}a estimation exhaustive, CR Acad.
  Sci. Paris 260~(1265-1266) (1935) 85.

\bibitem{Koopman-1936}
B.~O. Koopman, On distributions admitting a sufficient statistic, Transactions
  of the American Mathematical society 39~(3) (1936) 399--409.
\newblock \href {https://doi.org/10.2307/1989758} {\path{doi:10.2307/1989758}}.

\bibitem{Brody-2007}
D.~Brody, A note on exponential families of distributions, Journal of Physics
  A: Mathematical and Theoretical 40~(30) (2007) F691.
\newblock \href {https://doi.org/10.1088/1751-8113/40/30/F01}
  {\path{doi:10.1088/1751-8113/40/30/F01}}.

\bibitem{Cencov-1981}
N.~N. Cencov, Statistical decision rules and optimal inference, Transl. Math.
  Monographs, vol. 53, Amer. Math. Soc., Providence-RI (1981).

\bibitem{Campbell-1986}
L.~L. Campbell, An extended cencov characterization of the information metric,
  Proceedings of the American Mathematical Society 98~(1) (1986) 135--141.
\newblock \href {https://doi.org/10.1090/S0002-9939-1986-0848890-5}
  {\path{doi:10.1090/S0002-9939-1986-0848890-5}}.

\bibitem{Beck-2003}
C.~Beck, E.~G.~D. Cohen, Superstatistics, Physica A: Statistical mechanics and
  its applications 322 (2003) 267--275.
\newblock \href {https://doi.org/10.1016/S0378-4371(03)00019-0}
  {\path{doi:10.1016/S0378-4371(03)00019-0}}.

\bibitem{Kobayashi63}
S.~Kobayashi, K.~Nomizu, Foundations of Differential Geometry (Wiley Classics
  Library), Vol.~1, John Wiley and sons, New York, 1963.

\bibitem{Nawaz-2016}
S.~Nawaz, M.~Abedi, A.~Caticha, \href{arXiv.org/abs/1601.01708}{Entropic
  dynamics on curved spaces}, in: AIP Conference Proceedings, Vol. 1757, AIP
  Publishing LLC, 2016, p. 030004.
\newblock \href {https://doi.org/10.1063/1.4959053}
  {\path{doi:10.1063/1.4959053}}.
\newline\urlprefix\url{arXiv.org/abs/1601.01708}

\bibitem{Nelson-1985}
E.~Nelson, Quantum fluctuations, Princeton University Press, 1985.

\bibitem{pythonternary}
M.~{Harper}, et~al,
  \href{https://github.com/marcharper/python-ternary}{python-ternary: Ternary
  plots in python}.
\newblock \href {https://doi.org/10.5281/zenodo.594435}
  {\path{doi:10.5281/zenodo.594435}}.
\newline\urlprefix\url{https://github.com/marcharper/python-ternary}

\bibitem{Costa21}
F.~X. Costa, P.~Pessoa, Entropic dynamics of networks, Northeast Journal of
  Complex Systems 3~(1) (2021) 5.
\newblock \href {https://doi.org/10.22191/nejcs/vol3/iss1/5}
  {\path{doi:10.22191/nejcs/vol3/iss1/5}}.

\end{thebibliography}


\end{document}